\documentclass[sigchi]{acmart}

\AtBeginDocument{%
  \providecommand\BibTeX{{%
    \normalfont B\kern-0.5em{\scshape i\kern-0.25em b}\kern-0.8em\TeX}}}

\copyrightyear{2019}
\acmYear{2020}
\setcopyright{rightsretained}

\acmConference{}{}{}
\acmDOI{}
\acmISBN{}
\acmBooktitle{PREPRINT: UNDER REVIEW}

\begin{document}

\title{An Algorithmic Equity Toolkit for Technology Audits by Community Advocates and Activists\\(PREPRINT: UNDER REVIEW)}

\author{Michael Katell}
\affiliation{
  \institution{Information School\\University of Washington}
}
\email{mkatell@uw.edu}
\orcid{0000-0003-2200-6246}
\author{Meg Young}
\affiliation{%
  \institution{Information School\\University of Washington}
}
\email{megyoung@uw.edu}
\orcid{0000-0002-9300-8575}

\author{Bernease Herman}
\affiliation{%
  \institution{eScience Institute\\University of Washington}
}
\email{bernease@uw.edu}
\author{Dharma Dailey}
\affiliation{%
  \institution{Human Centered Design and Engineering\\University of Washington}
}
\email{ddailey@uw.edu}
\author{Aaron Tam}
\affiliation{%
  \institution{Evans School of Public Policy and Governance\\University of Washington}
}
\email{tama2@uw.edu}
\author{Vivian Guetler}
\affiliation{%
  \institution{Department of Sociology\\West Virginia University}
}
\email{vfg0002@mix.wvu.edu}
\orcid{0000-0001-5256-0104}

\author{Corinne Binz}
\affiliation{%
  \institution{Department of Computer Science\\Middlebury College}
}
\email{cbintz@middlebury.edu}
\orcid{0000-0001-9487-2069}
\author{Daniella Raz}
\affiliation{%
  \institution{School of Information\\University of Michigan}
}
\email{drraz@umich.edu}
\orcid{0000-0001-6259-9715}
\author{P. M. Krafft}
\affiliation{%
  \institution{Oxford Internet Institute\\University of Oxford}
}
\email{p.krafft@oii.ox.ac.uk}
\orcid{000-0001-8570-2180}
\renewcommand{\shortauthors}{Young, et al.}

\begin{abstract}
A wave of recent scholarship documenting the discriminatory harms of algorithmic systems has spurred widespread interest in algorithmic accountability and regulation. Yet effective accountability and regulation is stymied by a persistent lack of resources supporting public understanding of algorithms and artificial intelligence. Through interactions with a US-based civil rights organization and their coalition of community organizations, we identify a need for (i) heuristics that aid stakeholders in distinguishing between types of analytic and information systems in lay language, and (ii) risk assessment tools for such systems that begin by making algorithms more legible. The present work delivers a toolkit to achieve these aims. This paper both presents the Algorithmic Equity Toolkit (AEKit) Equity as an artifact, and details how our participatory process shaped its design.
Our work fits within HCI scholarship as a demonstration of the value of HCI methods and approaches to problems in the area of algorithmic transparency and accountability. 
%REVISIT: R1 highlights the question about how the toolkit addresses a question about whether it is enough for ADS to address bias issues, or whether facial recognition should be banned. The section explaining the design process now includes how this question emerged as an area of concern, how both the questionnaire and interactive demo were revised to highlight a "ban" as a substantive existing position in some jurisdictions.
\end{abstract}





\maketitle

\section{Introduction}
%TBH I like the level of detail in the following paragraph as an introduction
Extensive evidence demonstrates that the harms of algorithmic and information technologies are significant. Demonstrated harms exist across highly varied applications. Automated pretrial and sentencing risk assessment systems used in courts of law are racially biased \cite{desmarais2019PRETRIALRISKASSESSMENT, dressel2018AccuracyFairnessLimits, angwin2016MachineBiasThere}, facial recognition is racially and gender biased \cite{buolamwini2018GenderShadesIntersectional}, algorithmically supported hiring decisions are gender biased \cite{dastin2018AmazonScrapsSecret}, automated license plate readers lead to unwarranted police stops \cite{lecher2019PrivacyAdvocateHeld}, sensitive financial information has been stolen in major privacy breaches \cite{cowley2019EquifaxPayLeast}, digital currencies are susceptible to price manipulation \cite{gandal2018price,krafft2018experimental,donovan2019source}, social media is susceptible to disinformation campaigns \cite{woolley2018computational,tucker2018social}, labor platforms are exacerbating precarious labor conditions \cite{rosenblat2016AlgorithmicLaborInformation}, and much more.

%In an era of increasing technological advancement, our society now has more access than ever to information and data analysis tools, and the sphere of privacy has shrunk. Surveillance tools are increasingly used by government to track and monitor people, identifying who they are, who they know, what they do, and what they say. Automated decision system (ADS) technologies have been introduced to help organizations process, analyze, and interpret the plethora of data captured through surveillance tools. Even while these algorithmic tools are intended to help people make decisions in a seemingly objective manner, we join scholars and activists who contend that their use replicates, exacerbates, and masks inequities and injustice. 
%Extensive evidence demonstrates that the harms of surveillance and ADS technologies are significant. Surveillance systems like gunshot locators, transit microphones, deep packet sniffing, and stingrays may collect unsolicited conversations and private information. [cite ACLU] ADS technologies like pretrial and sentencing decision systems, facial recognition, and recruiting and hiring support systems have been found to have racial and gender bias (Buolamwini, J., & Gebru, T 2018). The increased use of ADS technologies has compromised public oversight because the public is less able to comprehend these technologies (Friedman, 1996, p. 342). Governments have rapidly expanded their use of surveillance and ADS technologies without considering the harms and biases they may hold.

Community organizations and civil rights groups concerned about the discriminatory risks of public sector technology adoption have pushed for algorithmic equity---including accountability, transparency, and fairness---through the implementation of municipal ordinances in several U.S. cities. These ordinances manage the acquisition and use of surveillance technologies and other automated decision systems. For instance, Berkeley, Cambridge, Nashville, Seattle, and others have passed surveillance ordinances to provide a degree of oversight for regulating such government technologies \cite{CommunityControlPolice}. 
Particular to the context of our research, the City of Seattle passed one of the first and strongest surveillance ordinances in 2017, mandating the publication of a ``master list'' of government surveillance technologies and a series of ``surveillance impact reports'' (SIR) that include input from both city personnel and designated community representatives \cite{gonzalez2017SeattleSurveillanceOrdinance,harrell2018SeattleSurveillanceOrdinance}. 

Yet existing legislation does not go far enough to address the risks at hand. Policy-makers and community members alike find algorithmic systems to be inscrutable and illegible. Risks that are already subject to existing legislation are not being recognized because the risks are tied to opaque and ill-understood algorithmic systems, and few existing pieces of passed legislation aim regulation at the algorithmic level. 
%As a result of these conceptual and legal gaps, public officials may not be paying close attention to the algorithmic features of their technical systems.
%Regulation specially tailored to algorithmic systems is needed in order to render automated decision systems accountable to the public.
Our prior work examined the Seattle Surveillance Ordinance as a case study and found that city personnel tasked with implementing that city's surveillance  ordinance did not consider any of the surveillance technologies in their portfolio to be algorithmic systems even when multiple technologies employed machine learning algorithms such as optical character recognition and facial recognition; these stakeholders instead focused on the technologies' data collection functions and privacy implications \cite{young2019municipal}. This finding suggests a crisis of legibility in algorithmic regulation. 
 %We identified a need for civil rights advocates to hold public officials accountable regarding their use of surveillance and ADS technologies, so we decided to develop a toolkit to achieve this goal. There are toolkits produced by AI Now, AI BlindSpot, and the Center for Government Excellence that go into great detail about how ADS technologies work and the potential areas where they may cause harm. However, these toolkits are designed for an academic, government, or technical audience and focus primarily on ADS technologies. Our 
 
%We respond to problems of legibility in public sector algorithmic systems by creating heuristic tools to help community organizers and engaged community members better identify surveillance and automated decision-making system (ADS) technologies and their risks. 
Our strategy to address the crisis of legibility around algorithmic systems is to empower community members to hold vendors and policy-makers accountable for these algorithmic harms that are being neglected.
Through interactions with a US-based civil rights organization and their coalition of community organizations, we identify a need for (i) heuristics that aid community organizers and community members in distinguishing between types of analytic and information systems in lay language, and (ii) risk assessment tools for such systems that begin by making algorithms more legible. The present work delivers a toolkit, the Algorithmic Equity Toolkit (AEKit) Equity, to achieve these aims. This paper both presents the AEKit as an artifact, and provides a case study about the participatory process that shaped its design.  We report on our iterative participatory design sessions with members of local civil rights organizations as the intended user base of the toolkit, and the Diverse Voices panels \cite{young2019InclusiveTechPolicy} we conducted to elicit additional input from those with lived experience of the harms motivating the toolkit.

%We respond to problems of legibility in public sector algorithmic systems by creating heuristic tools to help community organizers and engaged community members better identify surveillance and automated decision-making system (ADS) technologies and their risks. 
%In our participatory design sessions, community members reported that they would like to be better equipped to engage with public officials in their regulation.
%definitely need to make time to integrate stuff from Diverse Voices panels

\section{Related Work}
The community of human computer interaction (HCI) researchers is increasingly turning to the role that algorithms play in shaping sociotechnical systems, such as social media platforms, labor markets, and reputation scores. Algorithms are opaque due to multiple factors, including trade secret, technical unfamiliarity, and the complexity of machine learning and related techniques\cite{burrell2016machine}. As a result, users are sometimes not aware of the role algorithms play in shaping social worlds \cite{eslami2019user}. Indeed, researchers have highlighted the seeming ``invisibility'' of algorithms in sociotechnical systems. How embedded an algorithm is in a system changes with respect to the viewer; that is, algorithms are ``relational'' like attributes of other infrastructural systems \cite{star1999ethnography}. 
%Others describe this property as an algorithm's ``legibility'' to non-expert users \cite{young2019municipal}.  

Users form their own beliefs about how algorithms work \cite{rader2015understanding}, sometimes referred to as ``algorithmic literacy'' \cite{rainie2017theme}. These beliefs may not adhere closely to the way an algorithm works \cite{eslami2015always}. Nevertheless, these lay understandings, or ``folk theories'' of algorithms \cite{devito2018people,devito2018algorithm} shape user behavior \cite{nagy2015imagined}.  Advanced users try to leverage what they know about how a system works in order to achieve more visibility on social media feeds \cite{bucher2012want,cotter2019playing,bishop2019managing}. Existing approaches for making the functioning of algorithmic systems more transparent have primarily adopted the form of textual explanations; for example, of what, how, or why a newsfeed algorithm performed as it did \cite{rader2018explanations}; counterfactual explanations of what set of circumstances would result in a different algorithmic decision \cite{wachter_counterfactual_2017}; and explanations of how a particular personalized ad was shown to a specific user \cite{eslami_communicating_2018}. Some work has explored the potential for regulation to mandate such explanations \cite{selbst2017meaningful}. Amid growing interest in this area, researchers call for further application of methods found in HCI toward more user-centered design of these tools \cite{inkpen_where_2019}.

Our toolkit relates to and differs from existing related HCI efforts addressing this need. Compared with similar toolkit efforts such as the AI Blindpots toolkit \cite{AIBlindspotDiscovery}, we emphasize civically engaged community stakeholders as our intended users rather than companies or government agencies that design and deploy the algorithms. In that way, similar to Woodruff et al. \cite{woodruff2018qualitative}, we seek to empower historically marginalized communities. In contrast to the work of Woodruff et al., though, we do so through providing tools that enable recognition of and engagement with algorithms rather than soliciting perspectives on salient algorithms as they are readily understood. 
We also observed that other notable flowcharts that define and demystify AI for non-experts similarly rely on anthropomorphic metaphors, such as asking whether a system can ``see’'.\footnote{\url{https://www.technologyreview.com/s/612404/is-this-ai-we-drew-you-a-flowchart-to-work-it-out/}} Our toolkit attempts to avoid the use of these metaphors and adhere more closely to describing system functions. By trying to design it to be adaptable to a wide range of systems, the tool highlights that even conventional systems like Microsoft Excel could be used for automated decision system processes, and should be subject to oversight.
Finally, in comparison to methods in the area of explainable AI (e.g., \cite{miller2019explanation}) we develop a tool for first identifying the presence of an algorithm as a path to understanding it, a necessary prerequisite to pursuing explainability, and we emphasize consideration of sociotechnical context (cf., \cite{selbst2019fairness}).

%As Taina Bucher 

%Whereas the discussion of algorithmic transparency to date has largely focused on understanding the rationales adopted to achieve a particular assessment in a given system, there is a closely related question of when the use of algorithmic assessments in use are visible and to whom. 

\section{Design Context}

Our research takes place in a major U.S. city that has implemented a strong municipal surveillance ordinance. The state's legislature also recently drafted a tech fairness bill that is a first step in the direction of broad algorithmic regulation. Yet, previous research indicates that even expert policymakers are not prepared to understand the particular risks of algorithmic systems as such. In this participatory research project, we designed a toolkit that can be adopted within government, by civil rights organizations, and by individual community organizers to strengthen existing, ongoing, and future regulatory efforts.   

Our work intervenes in a critical gap in non-expert understanding of complex (and proprietary) algorithmic systems. Both within and beyond the public sector, grassroots and advocacy organizations desire visibility into systems that could have disparate impact on historically marginalized communities, but they lack domain knowledge and a set of recommended processes for exposing such systems to oversight. Furthermore, such systems are typically "black boxes," provided by vendors who are often unwilling to reveal key aspects of their functionality. Even when a system's functions are well-documented, the vectors of disparate impact are not readily apparent. To remedy this gap in understanding, and to provide those affected with tools necessary to hold algorithmic systems accountable, we co-designed the Algorithmic Equity Toolkit (AEKit) with community stakeholders in order to equip non-experts with a process and tools for surfacing unintended impacts of systems in use.
%In the study by Young et al. referenced in the introduction, the primary surveillance ordinance discussed mandates that city departments disclose the surveillance technologies they are using and subjects those technologies to political and community oversight. In a related move, proposed state legislation "requires the chief privacy officer to adopt rules regarding the development, procurement, and use of automated decision systems by a public agency". 
%Online research:
%We read op-ed news articles that called for banning facial recognition and the rationale behind this thinking. These articles also highlighted other government bodies across the U.S. that have already banned facial recognition, and they informed our understanding of why there are people that feel so strongly that facial recognition needs to be banned. 
\section{Methods}
%CRIBBED LANGUAGE ON PARTICIPATORY DESIGN
%THEN RESEARCH THROUGH DESIGN -- how we learned about the difficulties of joining expert and non expert views and creating a heuristic tool that holds across language

We iteratively developed the AEKit through a participatory design process that engaged data science experts, community partners, and policy advocates, while also drawing upon an array of prior literature \cite{dillahunt_reflections_2017, erete_intersectional_2018, green_data_2018} and similar toolkit efforts \cite{davidandersonEthicsAlgorithmsToolkit, AIBlindspotDiscovery}. % ADD CITES FROM BELOW
Initially, based on the regulatory focus of prior academic research, we envisioned that the primary users of the Algorithmic Equity Toolkit would be employees in state and local government seeking to surface the potential for algorithmic bias in existing systems. We thought advocacy and grassroots organizations could also find the toolkit useful for understanding the social justice implications of public sector systems. 
%A successful version of the toolkit would, for example, identify Automated License Plate Reader (ALPR) technology as an algorithmic system with known technical limitations (false positives) and potential for discriminatory impacts (pretexts to traffic stops). 
%(1) a technical questionnaire for identifying algorithmic systems and their attributes; (2) a stepwise evaluation procedure for surfacing the social context of a given system, its technical failure modes (i.e., potential for not working correctly, such as false positives), and its social failure modes (i.e. its potential for discrimination when working correctly). The toolkit will also include (3) a lightweight interactive tool for policymakers and non-experts that illustrates the relationship between how models are trained and adverse social impacts. 

%The project seeks to answer questions around algorithmic ethical issues and concerns by designing an Algorithmic Equity Toolkit.
Through our participatory design process, we refined our audience and design goals to focus on helping civil rights advocates and community activists---rather than state employees---identify and audit algorithmic systems embedded in public-sector technology, including surveillance technology. 
%This project will create an Algorithmic Equity Toolkit, a set of tools for identifying and auditing algorithmic processes used in the public sector, especially of surveillance and automated decision making technologies. The toolkit will include three primary components: 
We achieve this goal through three toolkit components:
(1) A flowchart designed for lay users for identifying algorithmic systems and their functions; (2) A Question Asking Tool (QAT) for surfacing the key issues of social and political concern for a given system. These tools together reveal a system's technical failure modes (i.e., potential for not working correctly, such as false positives), and its social failure modes (i.e. its potential for discrimination when working correctly); and (3)
An interactive web tool that illustrates the underlying mechanics of facial recognition systems, such as the relationship between how models are trained and adverse social impacts. In creating our own toolkit, we followed a weekly prototyping schedule interspersed with stakeholder feedback and co-design sessions.

The underlying questions that drove our design of these components were: What ethical issues should civil rights advocates be concerned with in regards to surveillance and automated decision systems? How are algorithmic systems reinforcing bias and discrimination? What do community organizers and non-tech experts understand about algorithmic tools and their impacts? What should they know about surveillance and automated decision systems to identify them and know how they work?
In comparison to existing resources, which tend to target software engineers and in some cases policymakers as an audience, we focused on policy advocates and community activists as users. 
%Toolkits that we reviewed included AI Now (www. https://ainowinstitute.org/), AI Blindspot (https://aiblindspot.media.mit.edu/), and Gov Ex Toolkit (https://govex.jhu.edu/wiki/center-for-government-excellence-releases-first-of-its-kind-algorithm-toolkit-to-reduce-bias-affecting-residents-from-automated-decisions-made-by-local-governments/). 

%We submitted new prototypes by the end of the work day on Fridays, received feedback early in the following work week, and then incorporated the feedback and new design ideas into the next prototype.

\subsection{Team Composition}

Our team consisted of a mix of students and researchers with expertise in policy analysis, qualitative research, human-centered design, computer science, data science, information ethics, and sociology.

\subsection{Stakeholders}

We envision our target users employing our toolkit to better inform their activism efforts in regards to tech fairness policy. We foresee community organizersand organization leaders using the toolkit to aid their understanding of the different functions of government technologies and the potential biases found in the use of algorithmic systems in their specific city and society at large.
Stakeholder engagement was a key component in the development of our toolkit. 
As the work began, the group articulated reasons for engaging directly with community stakeholders in our designn activities. These reasons included: (i) scoping and defining the problem space; (ii) understanding the broader context of the problem; (iii) testing and interpreting motivating concepts; (iv) assessing the usefulness and accessibility of the toolkit; (v) prototyping; (vi) ensuring technical accuracy; and (vii) planning for toolkit implementation and stewardship.

We partnered with a prominent civil right organization at the forefront in advocating for transparency and accountability from state and local governments, and two member organization from their network that advocate for the rights of historically marginalized communities. All of the stakeholders have collaborated together for tech fairness and advocacy work and have demonstrated enthusiasm for our toolkit to help their members raise concerns about the potential harms of algorithmic systems to policy makers and other public officials.

\subsection{Engagements}

%a methodology for surfacing the including the perspectives of members of social groups who are typically absent from the design emerging technologies and the policy making that responds to those technologies.
%A Diverse Voices panel consists of a small "experiential expert" heuristic evaluation session of individuals drawn from populations with relevant lived experiences who are often left out of design processes. We convened 3 panels each of 90 minutes---one of racial and social justice activists, one of immigrants, and one of formerly incarcerated people.
%Our panels included participation from race and social justice advocacy, immigrant experiences, and the formerly incarcerated.
%These participants provided feedback on the AEKit and offered insights into how surveillance and algorithmic technologies affect their communities. 

The team also met regularly with five members of a data science lab to receive feedback on the definitions and conceptualizations used in toolkit. 

\subsection{Ethics}

\subsubsection{Messaging}
Algorithmic harm is an issue reflective of systemic and structural inequality. When evaluating technologies used to manage and control populations, it is not uncommon to limit the discussion of merits to the scope of functionality; asking only if results are "accurate," "effective," or "predictive." Our concern is broader---through the design of this toolkit, we seek also to interrogate the social context of technology and to surface risks to the goals of establishing and maintaining a just and civil society. When confronted with facial recognition systems, for example, in addition to questioning their accuracy or to advocate for diversified data models to improve identification of women and people of color, we push farther and also question the whether this software should have a place in a democratic society at all; whether the potential benefits of a perfectly accurate facial recognition system outweigh the panoptic harms. 

\subsubsection{Stakeholders}
Our aim was to incorporate as much input as possible from our stakeholders without customizing the tool too much to the needs or desires of one specific group. One of our criticisms of existing tools is that they have not gone far enough to engage with stakeholder perspectives outside of academia and industry. As a response to this, we have chosen to focus heavily on the needs of underrepresented populations and members of historically marginalized communities in particular. However, given both the quantity and diversity of community stakeholders, creating a tool to service such a breadth of users presents a challenge. Thus, one ethical consideration is whether to design this tool with all, several, or one stakeholder(s) in mind. 

\subsubsection{How we address ethical concerns?}
Connecting with diverse stakeholders and a coalition of community groups, as well as co-designing with the a leading civil rights organization and other stakeholders to ensure our prototype and final product addresses stakeholders' concerns. While stakeholder engagement was fruitful and educational for the team, it also presented some constraints and limitations. Algorithmic systems are complex and are understood differently between lay and expert observers. Feedback about the toolkit content from community organizers differed starkly from that of the data scientists. The latter favored rich descriptions of the algorithmic processes that correctly identify machine learning concepts, such as clustering and classification. Meanwhile community stakeholders were easily overwhelmed by such technical language. This tension revealed the challenge of making an interpretive tools that is accessible to lay users while not misrepresenting the computational concepts they are designed to make salient. While balancing the competing desires of these stakeholder groups was difficult, it ultimately forced us to continually revise our approach in making the tool accessible while still remaining meaningful.

\section{Participatory Design}

In this section we describe how participants impacted the design at the outset of the project and throughout. These impacts are summarized in Table \ref{table:impacts}. The AEKit was initially formulated as an ultimate result of  initial conversations with our partnering civil rights organization.  In those initial conversations we had learned that the organization needed technical expert support on technology policy advocacy to provide additional input on technology, and for communicating why community engagement is needed as a robust part of policy-making efforts. In association with our responding to those needs, we identified the opportunity to develop pedagogical tools for algorithmic literacy. While initially we conceptualized these pedagogical tools as oriented towards a policy-maker audience, our partner civil rights organization encourage us to instead think about how we might empower community organizers and community members.  The project benefited greatly from a standing coalition of community organizations assembled by this civil rights organizations.  Members of this coalition became further partners in our participatory design process.  Through interactions with these community organizations, and with the Diverse Voices panels we convened, we further emphasized contexts of use of algorithmic technologies and simplicity and accessibilty of the toolkit.

\begin{table*}[]
\begin{tabular}{|p{4cm}|p{9cm}|}
\hline
\textbf{Time}              & \textbf{Impact}                                                                                                                                                                                                                                                                                                                                                                                                                                             \\ \hline
May 2018                   & In initial conversations with a partnering civil rights organization, we learned that the organization needed technical expert support on technology policy advocacy to provide additional input on technology, and communicating why community engagement is needed as a robust part of policy-making efforts.                                                                                                                                             \\ \hline
June 2018 - September 2018 & Initial research into technical capabilities of disclosed surveillance technologies and municipal oversight process emphasized need for an intervention on public understanding of algorithmic technologies                                                                                                                                                                                                                                                 \\ \hline
January 2019               & Initial design of Algorithmic Equity Toolkit proposed, targeting policy makers with a tool for identifying algorithmic systems and a checklist of red flags                                                                                                                                                                                                                                                                                                 \\ \hline
January - February 2019    & Partnering data science institute encourages inclusion of a technical component to the project to better suit learning objectives of students on the project. This technical component is formulated to be a web demo.                                                                                                                                                                                                                                      \\ \hline
February 2019              & Civil rights organization joins as a formal partner in the project                                                                                                                                                                                                                                                                                                                                                                                          \\ \hline
February 2019              & We held planning conversations with partners asking, "Given time and resource constraints, what process should our co-design follow?" Community partners requested continuous engagement in our process in addition to our initially planned Diverse Voices panels near the end. Based on this feedback, our team pivoted to a project that was participatory throughout.                                                                                   \\ \hline
June 2019                  & We held an initial meeting with our primary partner organization, asking "What does your advocacy look like in this space?" "Is there anything that would be useful to your organizing efforts?" After input from this partner, we pivoted to focus on supporting the organizing efforts of civil rights organizations through empowering community organizers and activists rather than targeting policymakers as our main audience in the first instance. \\ \hline
June 2019                  & We held initial envisioning sessions with small groups comprised of members of community partner organizations, asking "What is your current capacity to advocate in the area of algorithmic decision systems?" "What would support your work?" Learning about the policy context led us to focus on intervening at the public comment period of oversight surveillance and ADS technologies.                                                               \\ \hline
July 2019                  & We held another round of co-design sessions with participants sharing initial artifacts. Participants directed the toolkit design to be less technical to enable broader diffusion and use. As a result, the toolkit shifted from a focus on machine learning concepts to embracing the wider sociotechnical context of use (e.g. "Is the operator being trained in the accuracy levels of the system?)                                                     \\ \hline
August 2019                & We conducted 3 Diverse Voices panels with members of communities historically harmed by surveillance and ADS technologies. Panelists identified several accessibility barriers; as a result of this input we modified the toolkit design to be more concise for field use, and more  focused on algorithmic harm.                                                                                                                                           \\ \hline
\end{tabular}
\caption{A timeline of major impacts of participatory engagements.}
\label{table:impacts}
\end{table*}

\section{The Algorithmic Equity Toolkit}

\begin{figure}
    \centering
     \includegraphics[width=0.7\columnwidth]{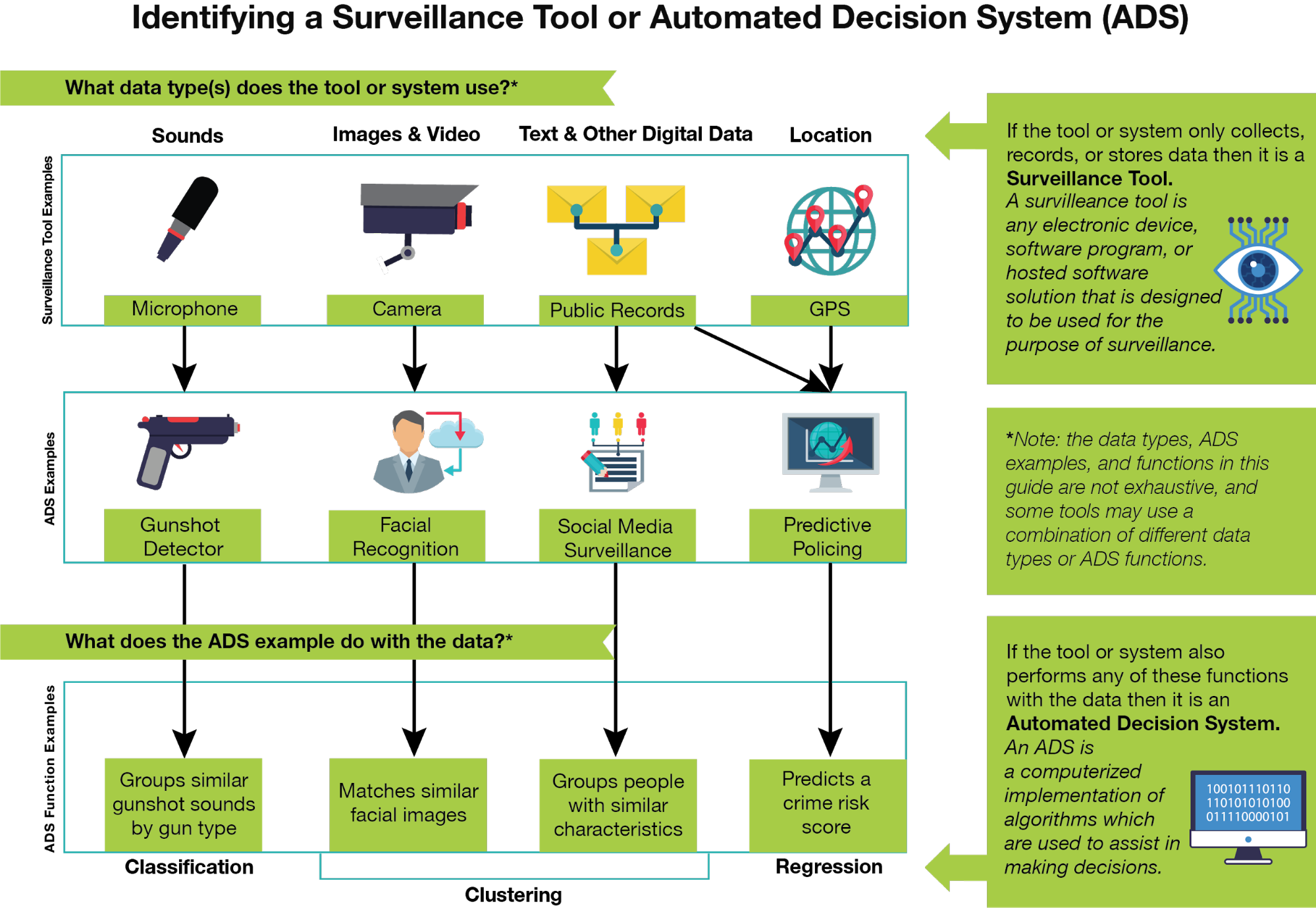}
     \includegraphics[width=0.7\columnwidth]{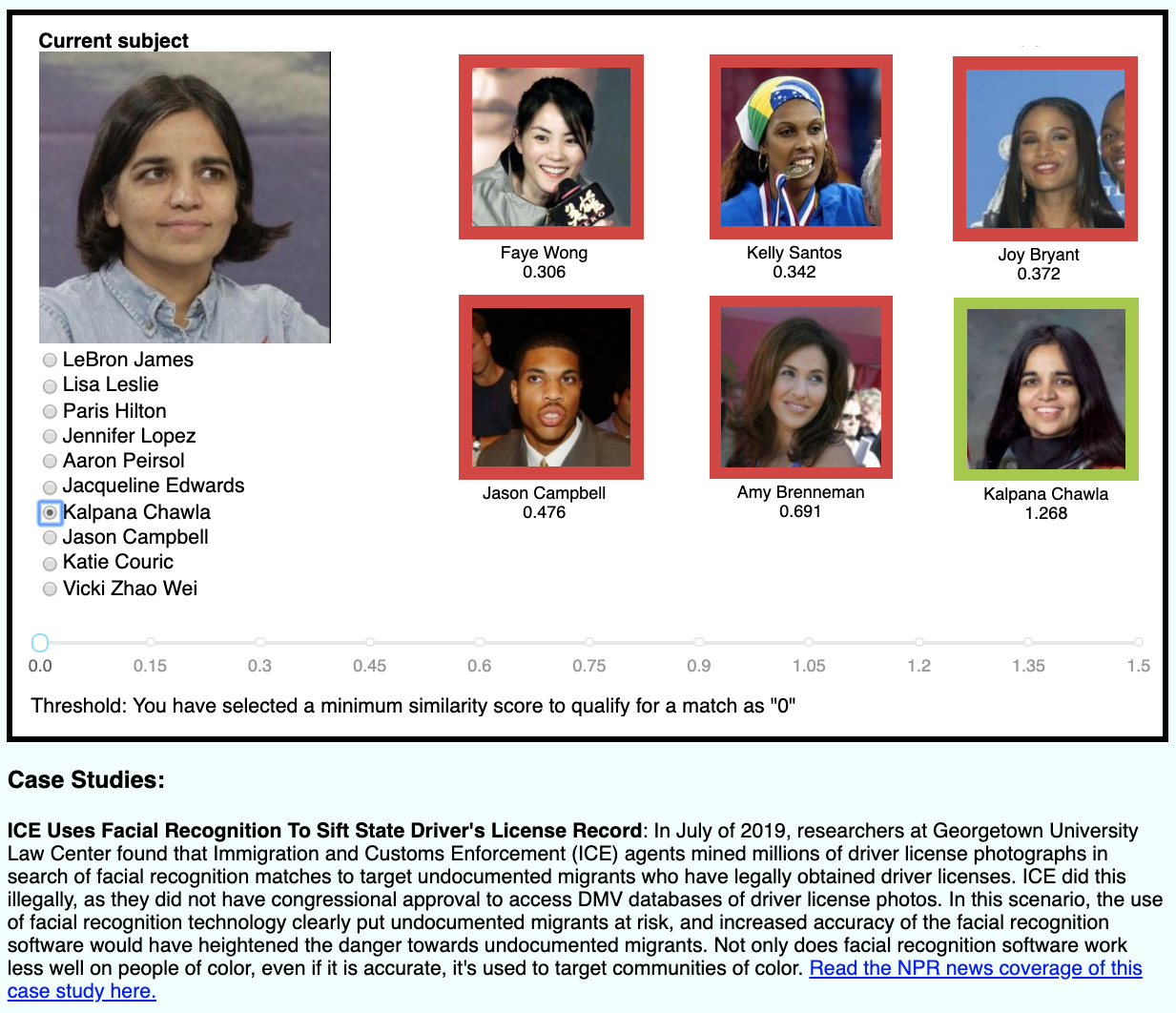}
     \includegraphics[width=0.7\columnwidth]{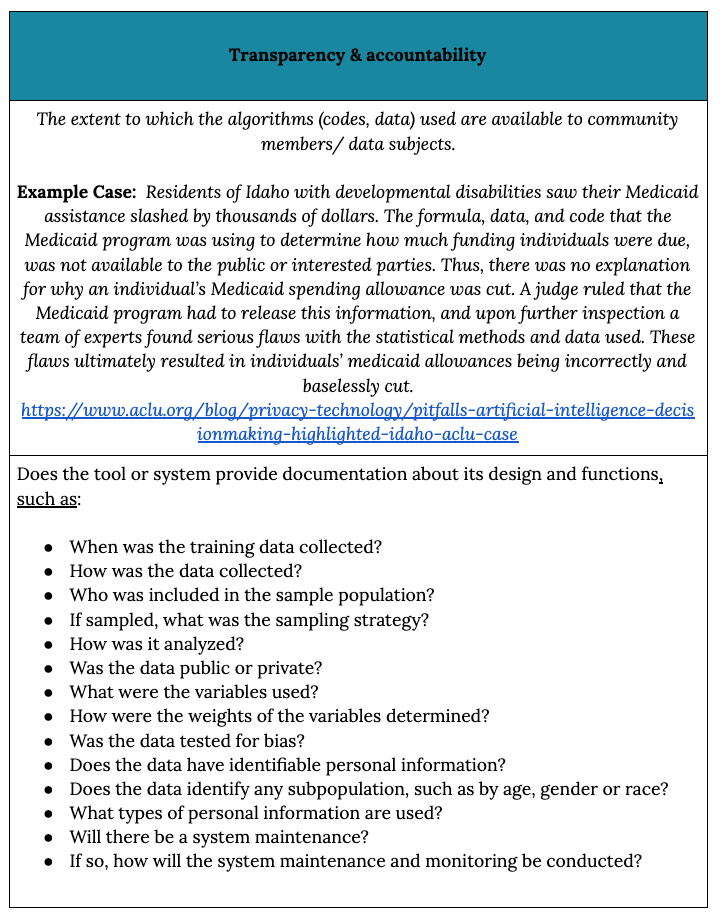}
    \caption{The three primary components of the Algorithmic Equity Toolkit, consisting of (a) an identification guide for automated decision systems, (b) a questionnaire on potential harms, and (c) an interactive tool demonstrating algorithmic bias for a particular technology. In its initial scope, we conceived the primary users of the Algorithmic Equity Toolkit as employees in state and local government seeking to surface the potential for algorithmic bias in existing systems. The choice to center advocacy and grassroots organizations ---particularly in support of their policy advocacy work in this space-- emerged from our participatory design process early on.}
    \label{fig:my_label}
\end{figure}

At the time of writing, the toolkit has three components:\footnote{The latest version is available online here: \url{https://github.com/anon770/toolkit_anon}. The repository is organized to demonstrate how the toolkit evolved in response to stakeholder engagements.}
\begin{enumerate}
\item A flowchart for distinguishing surveillance and ADS's and their different functions. 
\item A question-asking tool for surfacing the social context of a given system, its technical failure modes (i.e., potential for not working correctly, such as false positives), and its social failure modes (i.e. its potential for discrimination when working correctly). 
\item An interactive demo of facial recognition that reveals the underlying harms and mechanics of facial recognition technology. 
\end{enumerate}
Illustrations of the versions of these three components at the time of writing are shown in Figure \ref{fig:my_label}.

The primary users of the Algorithmic Equity Toolkit will be community organizers, including civil rights advocacy and grassroots organizations as well as anyone interested in algorithmic equity. A secondary target audience includes personnel tasked with implementing technologies for managing and controlling populations. A key goal is to overcome power asymmetries between individuals and systems of authority, such as government agencies who should be held accountable for the technologies they implement in their communities. The toolkit can be used when engaging with policymakers and other public officials, or in other contexts where individuals and groups want to learn more about surveillance and ADS technologies and their potential harms.

An ADS is a computerized implementation of an algorithmic system to assist in decision-making by humans, or to take specified actions automatically. ADS's are increasingly used in our society to analyze data and make decisions more quickly and efficiently; however, the increasing use of ADS's decreases transparency and accountability due to their complexity and the lack of public awareness about how they work. 
The steps for using our present toolkit are:\\

\begin{itemize}
    \item[Step 1:] Start with the Surveillance and ADS Identification Guide.
This guide should be used to help you determine whether a government technology is a surveillance or ADS tool or system. It will also help you understand the different functions of surveillance and ADS tools and systems. With this Surveillance and ADS ID Guide, civil rights advocates can better detect the presence of algorithms and what those features do. \item[Step 2:] Questionnaire.
Use the questionnaire to inquire about the potential harms of surveillance or ADS technologies when engaging with policymakers and other public officials.
\item[Step 3:] Interactive facial recognition web demo.
Click on [link]  to access the interactive demo on facial recognition tool that illustrates some of the harms of the technology.
\end{itemize}

\subsection{Flowchart for identifying a machine learning or AI system}

% I question if we need to include an 'unmet need' section for each element of the tool, and instead make our case or the entire toolkit meeting an unmet need. This will reduce word count and increase concision IMHO (mk)
\subsubsection{Unmet need:}
Information technologies are an increasing part of our everyday lives. Some technologies are more impactful than others, potentially affecting individual and group autonomy, civil rights, and safety. Our work with community groups and civil rights activists suggests that a means of ensuring that the effects of information technologies are mainly positive, or that their negative aspects are minimized, begins at recognizing and understanding the technologies in our midst. This is particularly true of public-sector technologies, where the principles of democratic governance require that state actors be accountable to the public for the tools and technologies they use to manage and control the population. Research by Young et al. \cite{young2019municipal} suggests that the public, including policy makers, need assistance in identifying the opaque algorithmic aspects of public sector systems so that technology implementations can be sufficiently transparent and publicly accountable. 

\subsubsection{Meeting the need of helping community organizers understand: Where is the algorithm in this system---what is the algorithm doing?} 
As described by Young et al., lay observers, including professionals who should know, often do not recognize that a system is "algorithmic". At other times, people may know a technology is algorithmic, but they don't know how the algorithm is coming into play. In still more cases, there are systems that can be understood as algorithmic but their harms are not necessarily of concern (e.g. simple calculators, thermostats). The goal of the flowchart tool is to signal the likely presence of algorithms that likely pose harms, especially harms that correspond to marginalized identities and histories of discriminatory state action. The tool represents a set of definitional criteria, which, when applied to algorithmic systems, help to scope which technologies should be part of the conversation. 
%How do we develop basic questions that identify those and not others?

\subsubsection{Form:}

The tool we developed guides users through a process for identifying components of technical systems that are algorithmic. Many technological artifacts are ambiguous as to their inner functionality leaving observers, including users, unaware of what kind of work the artifact does over and above its most obvious functions. To make the embedded features more salient and open to questioning, our flowchart tool offers a decision tree for contemplating what has been disclosed or can be observed about a technology, providing a verdict about whether it might be an AI system. While some systems are relatively straightforward, either because their functions are obvious, publicized, or fully disclosed, there are other technologies that are more challenging to unpack. An example of the former is booking photo comparison software (BPCS), which employs an algorithmic system that has already faced considerable public scrutiny, facial recognition. Many other artifacts contain algorithmic features that are much harder to detect simply by encountering them or even by having them explained by a public official or software vendor. 

The flowchart differentiates algorithmically-enhanced systems from systems that are merely surveillant (i.e. only a data collection tool and not a tool that performs, say, an analysis and/or renders action-guiding judgements, or takes its own actions). An automated license plate reader (ALPR) may appear at first to be merely surveillant---basically a device that captures license plate images. But embedded within are AI components such as computer vision and algorithms for recognizing alpha-numeric sequences and matching the results to lists of license plates of interest. It is helpful to understand these features because, over and above whatever functionality is most obvious (e.g. a camera), embedded systems have their own failure modes, design constraints, and social valences that can contribute to the artifact's impact on individuals and communities. For example, some ALPR systems do not detect the issuing state of a license plate suggesting that a driver from Arizona could be misidentified as a driver from Pennsylvania whose license plate contains a similar alpha-numeric sequence. Even when such a system accurately identifies a license plate of interest, there are questions about the social conditions that lead to drivers being subjects of detection, such as the correlation between unpaid parking tickets and racialized poverty, that cannot be asked without peeling back the layers of technology to the sociotechnical imaginaries bundled within.

%We looked to examples produced by others for inspiration. For example, we were impressed by Examples for inspiration: 
%AI flowchart
%Algorithmic Cheatsheet
%Surveillance Questionnaire
%Interface: The flowchart must be intuitive and legible to non-technical users. 
%Flowchart (1-2 page) PDF format. Optional interactive digital interface.
%Questions should employ familiar language to the extent possible (e.g. Does the system take photos and pull information out of them?)
%The use of technical terms should be minimized and/or accompanied by definitions or explanations
%Potential features:
%Include worked out examples of 3-5 technologies (a couple obvious ones that qualify and don't, plus a couple harder ones). See e.g. the technologies we previously used on AI survey.
%Once a user completes the process, the user receives a verdict as to the algorithmic aspect of a system, and potentially a brief description of the technologies likely present in the system. 
%The result could also include which classes of algorithms are likely to be used (e.g. This system appears to use machine learning.)

\subsection{Asking the right questions}

\subsubsection{Unmet need:}
Having identified an algorithmic system, the next step is to pose questions about it; about its functions and features, about the claims made about its efficacy, and about its potential to harm those to whom it is applied. Armed with a narrowly tailored set of questions, community organizers and activists can contest the narratives provided to them by authority figures and product vendors, proposing richer shared meanings onto the technologies in question.  
Given a camera with facial recognition capabilities, for example, Toolkit users will be able to address concerns about this technology, such as issues of race and gender detection parity and the potential for the tool contribute to oppressive feedback loops in which systemic discrimination is reproduced through the use of the tool by institutions with a history of discriminatory action. 
In creating this tool, we set some baseline standards, including: (i) it must be intuitive and legible to non-technical users; and (ii)
questions should employ familiar language to the extent possible. %(See Appendix ___ )

\subsubsection{Form:} The Question Asking Tool (QAT) is a tool for guiding users through the salient issues presented by an algorithmic system. Its goal is to surface social contexts and technical failure modes and to prompt questions that reveal potential harms, particularly harms to particular communities and identities. The QAT could also contribute to algorithmic impact assessments required by local and international laws (e.g. the General Data Protection Regulation) and recommended by legal experts and other scholars \cite{article29dataprotectionworkingparty2018GuidelinesAutomatedIndividual, koops2016TypologyPrivacy, edwards2017SlaveAlgorithmWhy}, including the public accountability processes required by municipal surveillance ordinances in the United States. The tool can also be used by individual community members in dialogue with public officials and other authority figures. The tool distills known harms from the Fairness, Accountability, and Transparency literature and translates them for non-specialist audience.

The QAT prompts toolkit users to identify the socio-ethical issues community advocates and civil rights activists should be concerned with in regards to algorithmic systems. In what ways does a particular type of algorithmic system reinforce bias and discrimination? What should individuals and groups with little or no technical expertise understand about the impacts of algorithmic tools? What answers should they demand from public officials and other authority figures implementing management and control technologies in their communities?
The QAT contains a series of questions sorted into categories designed to assess an algorithmic system's potential harms in regard to social impact, appropriate use, transparency and accountability, data security and privacy, and interpretability or operability. 

\subsection{Interactive demo of intersectional failures of facial recognition}

\subsubsection{Unmet need:}
Observers may have heard that algorithmic systems are problematic but may have difficulty envisioning and internalizing what those problems are. The interactive demo makes at least some issues of algorithmic sorting and decision making salient to the user. 

\subsubsection{Form:}

The interactive demo tool demonstrates the problem of algorithmic harms such as bias in machine learning due to technical limitations and model representation, among other problems. 
Our demo  involved running ten celebrity photos in Open Face's model using a database of 60 celebrity photos collected from Labeled Faces in the Wild and Google image searches. We then selected the top 8 closest images for each of the ten celebrity photos to include in our demo. Of all the ten celebrity photos, the minimum similarity score of the top 8 closest images was 0.15, between a photo of Aaron Peirsol and Ai Sugiyama, and the maximum similarity score was 1.384, between two different photos of LeBron James. Overall, celebrities with lighter skin tones had lower similarity scores than celebrities with darker skin tones.
Our demo showing differences in similarity scores along the lines of skin tone are consistent with the literature surrounding facial recognition software and accuracy according to skin tone  \cite{buolamwini2018GenderShadesIntersectional}.

\section{Discussion}
%BEGIN Dharma's text
We observed that our efforts toward equity in public-sector algorithmic systems required articulation work, or alignment, \cite{corbin1993articulation} between the expertise of three distinct groups: civil rights legal experts, technology experts, and those with the lived experience of being differentially targeted by ADS groups. The shortest path to integrating these different knowledges was by traversing the social distance between them with a prototype in hand, letting each stakeholder interaction inform our subsequent encounters. Through frequent, concurrent probing with each of these groups, the territory of the intervention space began to reveal itself. Though we aim for the Toolkit to serve as an education aid, reinforcing connections between these three critical groups was no less important to us. The former is the foundation for individual awareness. The latter is the foundation for the collective action needed to propel tactical and just action-- that can move ADS use closer toward social equity and accountability.  To effect such change, it is not enough for those impacted by algorithmic systems to better understand the mechanics of these technologies. They must have a sense of the recourse that may mitigate harms. 
%and increase benefits (including the ability to leverage existing legal frameworks). 

But pushing knowledge in one direction is not enough (c.f. the failures of the ``deficit model'' of public understanding of science \cite{sturgis2004science}). True change also demands technologists better understand the cultural, social and legal frames of these technologies as well as the lived experience of those particularly impacted by their designs. Likewise, legal and political experts better align the aims of civil society when they have a more grounded understanding of the technologies along with the  lived experience of the affected. Such multi-directional co-learning necessitates a more demanding design process in which the problem and potential solutions are articulated with by each respective expert. Initially this results in confusion and ambiguity as the ways of conceiving of these technologies was not mutually intelligible. After several iterations of articulation (and rearticulation), a solution space can emerge that is truly reflective of all these expertises. This co-produced understanding may be the most important contribution of this work. Yet the social and technological complexities of algorithmic technologies inevitably slow the progress of multilateral co-production. Our initial co-articulations are incomplete and provisional. We assess that it will take many years of such effort to achieve a fully articulated mutual understandable operational vision of Algorthimic Justice.  This work is but one early starting point. For this reason, we reflect on this work as an example of Research through Design \cite{bardzell2015ImmodestProposalsResearch, bardzell2016DocumentingResearchDesign, zimmerman2007ResearchDesignMethoda, zimmermanAnalysisCritiqueResearch, gaver2012WhatShouldWe}.
%whereby we place greater merit on what is being learned through the process of designing than we place on the artifacts we have produced so far. 
%END Dharma's text

%ACLU, WA as the primary stakeholder has provided connections to other community organizations such as Densho, CAIR who are members of the Tech Fairness Coalition. The organizations have provided insights, feedback and suggestions on the toolkit design and how to make it accessible for non-technical community members. The organizations have expressed interest in using the toolkit and sharing it on their websites and social media accounts.

%What impact has your project had, or do you anticipate it having?

%The primary goal of this project is to empower community members with a toolkit that helps them ask questions about algorithmic technologies and biases to their elected officials. In addition, we hope the toolkit will help inform local and national technology policy changes and lead to algorithmic equity.

%The City of Seattle and Washington State are both world leaders in technology policy. The Washington State House has drafted a tech fairness bill (HB 1655) a first step in the direction of broad algorithmic regulation. However, previous research indicates that even expert policymakers are not prepared to understand the particular risks of algorithmic systems as such. We anticipate the toolkit to be adopted both within government and by policy advocates such as the ACLU to strengthen HB 1655 and other existing, ongoing, and future regulatory efforts. 

\subsection{Limitations}
The AEKit has several limitations. First, we found that there exists little consensus about the definition and structure of concepts like artificial intelligence, machine learning, and automated decision systems. Second, meeting with many diverse stakeholders presented a challenge in building the toolkit. Such a diverse set of community organizers inevitably resulted in diverse and sometimes conflicting priorities, and it proved difficult to meet all expectations.
%Due to time constraints, we also could not possibly meet with all stakeholders in the community coalition. Thus, we had to select some community organizations over others resulting in certain community voices being more dominant and influential in creating our toolkit. 
We also faced a challenge of balancing unnecessary technical detail with simplicity.
Understanding the baseline level of knowledge of the user and what would be helpful and not helpful for them to know, and
connecting the flowchart to the checklist and interactive demo in a fluid way were also challenges. 
Finally, we did not want our toolkit to communicate that the harms we covered were  comprehensive. For instance, while the facial recognition demo showcases inaccuracies in facial recognition, it runs the risk of communicating to users that our goal is  accuracy in facial recognition. A more complete demo would attend more fully to the distinct and troubling harms of fully accurate face recognition and surveillance. To attempt to mitigate this risk, we decided to include case studies and quotes from stakeholders voicing this concern alongside the technical demo. In unpacking this tension, we also note that this failure may be inherent to the framing of fairness with respect to different social and demographic groups. As Hoffmann explains \cite{hoffmann2019fairness}, the hierarchical logic underpinning the discourse of ``fairness'' may reproduce disadvantage rather than mitigate it.

%This project faced several limitations. First, we found little consensus regarding the definition and structure of concepts like artificial intelligence, machine learning, and automated decision systems (Krafft et al. 2019). Second, meeting with many diverse stakeholders presented a challenge in building the toolkit. Such a diverse set of community members inevitably resulted in diverse and sometimes conflicting priorities, and it proved difficult to meet all expectations. Due to time constraints, we also could not possibly meet with all stakeholders in the community coalition; thus, we had to select some community organizations over others resulting in certain community voices being more dominant and influential in creating our toolkit.

%\subsection{Challenges}

%Purpose: To help civil rights advocates, organizers, and grasstop leaders identify what is an ADS and the different ways they work.

%Challenges: 

%I propose making the interactive demo the intermediary between the flowchart and checklist.
%Definition of an ADS

%\subsection{Learning from Other Areas}

A broader concern is the negative impact of the existence of flowcharts and checklists for accountability and regulatory work.
%Additionally, I was able to have a serendipitous conversation with one of the DSSG project leads who, as an architect and planning board member long active in affordable housing policy (3 decades). I asked him what he thought of neg decs as a model for public oversight in other areas where exchange was needed between expert and non-expert to achieve public oversight. Without hesitation he pointed out that 
In a better developed area of environmental regulation, the environmental review process and negative declarations are well intended and effective in many ways. However, that process has had unintended consequences that were not foreseen by those who designed and adopted the (now standard) environmental review processes.  For example, environmental reviews are used very effectively for class war with wealthier communities and individuals able to use effectively slow or stop any kind of development they find undesirable. This has had a tremendous and deleterious impact on affordable housing. The problem of hijacking environmental review for other ends is currently unfixable at this point (4 to 5 decades on this as well established procedure). One idea to resolve this kind of issue is potentially to have an independent review for both risks AND benefits.

%The larger point though is that, even as the team looks for models that can be appropriated from other places where expert knowledge requires specific means to attain public oversight, those models may come with warts. How to avoid those warts, seems like it probably fits more into future work than a critical path for the immediate next few weeks, but perhaps it warrants closer examination in future. 

\section{Conclusion}

Community organizers and civil rights activists throughout the U.S. are concerned about surveillance technologies being implemented in their communities. There is concern that these technologies are being used by law enforcement and other public officials for profiling and targeting historically marginalized communities. Activists and advocates have pushed for algorithmic equity (accountability, transparency, fairness) through the implementation of legislation like municipal surveillance ordinances that regulate and supervise the acquisition and use of surveillance technology. Major cities, including Seattle, Berkeley, Nashville, Cambridge, and others have implemented ordinances that differ in their scope, process, and power in regulating government technologies. However, most technology policy legislation in the U.S. fails to manage the growing use of automated decision systems such as facial recognition and predictive policing algorithms. 
Despite its limitations, the Algorithmic Equity Toolkit is a vital tool that community civil rights advocates can use to voice their concerns about these technologies during the decision-making process for the acquisition of these technologies.  Our work fits within HCI scholarship as a demonstration of the value of HCI methods and approaches to problems in the area of algorithmic transparency and accountability. 

\begin{acks}
Blinded for peer review.
\end{acks}

\bibliographystyle{ACM-Reference-Format}
\bibliography{main}


\begin{thebibliography}{54}

%%% the Web address.

\ifx \showCODEN    \undefined \def \showCODEN     #1{\unskip}     \fi
\ifx \showDOI      \undefined \def \showDOI       #1{#1}\fi
\ifx \showISBNx    \undefined \def \showISBNx     #1{\unskip}     \fi
\ifx \showISBNxiii \undefined \def \showISBNxiii  #1{\unskip}     \fi
\ifx \showISSN     \undefined \def \showISSN      #1{\unskip}     \fi
\ifx \showLCCN     \undefined \def \showLCCN      #1{\unskip}     \fi
\ifx \shownote     \undefined \def \shownote      #1{#1}          \fi
\ifx \showarticletitle \undefined \def \showarticletitle #1{#1}   \fi
\ifx \showURL      \undefined \def \showURL       {\relax}        \fi
% invisible to TeX
\providecommand\bibfield[2]{#2}
\providecommand\bibinfo[2]{#2}
\providecommand\natexlab[1]{#1}
\providecommand\showeprint[2][]{arXiv:#2}

\bibitem[\protect\citeauthoryear{??}{AIB}{[n.d.]}]%
        {AIBlindspotDiscovery}
 \bibinfo{year}{[n.d.]}\natexlab{}.
\newblock \bibinfo{title}{{{AI Blindspot}}: {{A Discovery Process}} for
  Preventing, Detecting, and Mitigating Bias in {{AI}} Systems}.
\newblock \bibinfo{howpublished}{https://aiblindspot.media.mit.edu/}.
\newblock


\bibitem[\protect\citeauthoryear{??}{Com}{[n.d.]}]%
        {CommunityControlPolice}
 \bibinfo{year}{[n.d.]}\natexlab{}.
\newblock \bibinfo{title}{Community {{Control Over Police Surveillance}}}.
\newblock
  \bibinfo{howpublished}{https://www.aclu.org/issues/privacy-technology/surveillance-technologies/community-control-over-police-surveillance}.
\newblock


\bibitem[\protect\citeauthoryear{Angwin, Larson, Mattu, and Kirchner}{Angwin
  et~al\mbox{.}}{2016}]%
        {angwin2016MachineBiasThere}
\bibfield{author}{\bibinfo{person}{Julia Angwin}, \bibinfo{person}{Jeff
  Larson}, \bibinfo{person}{Surya Mattu}, {and} \bibinfo{person}{Lauren
  Kirchner}.} \bibinfo{year}{2016}\natexlab{}.
\newblock \showarticletitle{Machine {{Bias}}: {{There}}'s {{Software Used
  Across}} the {{Country}} to {{Predict Future Criminals}}. {{And}} It's
  {{Biased Against Blacks}}.}
\newblock \bibinfo{journal}{\emph{ProPublica}} (\bibinfo{year}{2016}).
\newblock


\bibitem[\protect\citeauthoryear{{Article 29 Data Protection Working
  Party}}{{Article 29 Data Protection Working Party}}{2018}]%
        {article29dataprotectionworkingparty2018GuidelinesAutomatedIndividual}
\bibfield{author}{\bibinfo{person}{{Article 29 Data Protection Working
  Party}}.} \bibinfo{year}{2018}\natexlab{}.
\newblock \bibinfo{booktitle}{\emph{Guidelines on {{Automated}} Individual
  Decision-Making and {{Profiling}} for the Purposes of {{Regulation}} 2016/679
  (Wp251rev.01)}}.
\newblock \bibinfo{type}{{T}echnical {R}eport} WP251rev.01.
\newblock


\bibitem[\protect\citeauthoryear{Bardzell, Bardzell, Dalsgaard, Gross, and
  Halskov}{Bardzell et~al\mbox{.}}{2016}]%
        {bardzell2016DocumentingResearchDesign}
\bibfield{author}{\bibinfo{person}{Jeffrey Bardzell}, \bibinfo{person}{Shaowen
  Bardzell}, \bibinfo{person}{Peter Dalsgaard}, \bibinfo{person}{Shad Gross},
  {and} \bibinfo{person}{Kim Halskov}.} \bibinfo{year}{2016}\natexlab{}.
\newblock \showarticletitle{Documenting the {{Research Through Design
  Process}}}. In \bibinfo{booktitle}{\emph{Proceedings of the 2016 {{ACM
  Conference}} on {{Designing Interactive Systems}} - {{DIS}} '16}}.
  \bibinfo{publisher}{{ACM Press}}, \bibinfo{address}{{Brisbane, QLD,
  Australia}}, \bibinfo{pages}{96--107}.
\newblock


\bibitem[\protect\citeauthoryear{Bardzell, Bardzell, and
  Koefoed~Hansen}{Bardzell et~al\mbox{.}}{2015}]%
        {bardzell2015ImmodestProposalsResearch}
\bibfield{author}{\bibinfo{person}{Jeffrey Bardzell}, \bibinfo{person}{Shaowen
  Bardzell}, {and} \bibinfo{person}{Lone Koefoed~Hansen}.}
  \bibinfo{year}{2015}\natexlab{}.
\newblock \showarticletitle{Immodest {{Proposals}}: {{Research Through Design}}
  and {{Knowledge}}}. In \bibinfo{booktitle}{\emph{Proceedings of the 33rd
  {{Annual ACM Conference}} on {{Human Factors}} in {{Computing Systems}} -
  {{CHI}} '15}}. \bibinfo{publisher}{{ACM Press}}, \bibinfo{address}{{Seoul,
  Republic of Korea}}, \bibinfo{pages}{2093--2102}.
\newblock


\bibitem[\protect\citeauthoryear{Bishop}{Bishop}{2019}]%
        {bishop2019managing}
\bibfield{author}{\bibinfo{person}{Sophie Bishop}.}
  \bibinfo{year}{2019}\natexlab{}.
\newblock \showarticletitle{Managing visibility on YouTube through algorithmic
  gossip}.
\newblock \bibinfo{journal}{\emph{New Media \& Society}}
  (\bibinfo{year}{2019}), \bibinfo{pages}{1461444819854731}.
\newblock


\bibitem[\protect\citeauthoryear{Bucher}{Bucher}{2012}]%
        {bucher2012want}
\bibfield{author}{\bibinfo{person}{Taina Bucher}.}
  \bibinfo{year}{2012}\natexlab{}.
\newblock \showarticletitle{Want to be on the top? Algorithmic power and the
  threat of invisibility on Facebook}.
\newblock \bibinfo{journal}{\emph{New media \& society}} \bibinfo{volume}{14},
  \bibinfo{number}{7} (\bibinfo{year}{2012}), \bibinfo{pages}{1164--1180}.
\newblock


\bibitem[\protect\citeauthoryear{Buolamwini and Gebru}{Buolamwini and
  Gebru}{2018}]%
        {buolamwini2018GenderShadesIntersectional}
\bibfield{author}{\bibinfo{person}{Joy Buolamwini} {and}
  \bibinfo{person}{Timnit Gebru}.} \bibinfo{year}{2018}\natexlab{}.
\newblock \showarticletitle{Gender {{Shades}}: {{Intersectional Accuracy
  Disparities}} in {{Commercial Gender Classification}}}. In
  \bibinfo{booktitle}{\emph{Proceedings of {{Machine Learning Research}}}},
  Vol.~\bibinfo{volume}{81}. \bibinfo{address}{{New York, NY}},
  \bibinfo{pages}{15}.
\newblock


\bibitem[\protect\citeauthoryear{Burrell}{Burrell}{2016}]%
        {burrell2016machine}
\bibfield{author}{\bibinfo{person}{Jenna Burrell}.}
  \bibinfo{year}{2016}\natexlab{}.
\newblock \showarticletitle{How the machine ‘thinks’: Understanding opacity
  in machine learning algorithms}.
\newblock \bibinfo{journal}{\emph{Big Data \& Society}} \bibinfo{volume}{3},
  \bibinfo{number}{1} (\bibinfo{year}{2016}),
  \bibinfo{pages}{2053951715622512}.
\newblock


\bibitem[\protect\citeauthoryear{Corbin and Strauss}{Corbin and
  Strauss}{1993}]%
        {corbin1993articulation}
\bibfield{author}{\bibinfo{person}{Juliet~M Corbin} {and}
  \bibinfo{person}{Anselm~L Strauss}.} \bibinfo{year}{1993}\natexlab{}.
\newblock \showarticletitle{The articulation of work through interaction}.
\newblock \bibinfo{journal}{\emph{The sociological quarterly}}
  \bibinfo{volume}{34}, \bibinfo{number}{1} (\bibinfo{year}{1993}),
  \bibinfo{pages}{71--83}.
\newblock


\bibitem[\protect\citeauthoryear{Cotter}{Cotter}{2019}]%
        {cotter2019playing}
\bibfield{author}{\bibinfo{person}{Kelley Cotter}.}
  \bibinfo{year}{2019}\natexlab{}.
\newblock \showarticletitle{Playing the visibility game: How digital
  influencers and algorithms negotiate influence on Instagram}.
\newblock \bibinfo{journal}{\emph{New Media \& Society}} \bibinfo{volume}{21},
  \bibinfo{number}{4} (\bibinfo{year}{2019}), \bibinfo{pages}{895--913}.
\newblock


\bibitem[\protect\citeauthoryear{Cowley}{Cowley}{2019}]%
        {cowley2019EquifaxPayLeast}
\bibfield{author}{\bibinfo{person}{Stacy Cowley}.}
  \bibinfo{year}{2019}\natexlab{}.
\newblock \showarticletitle{Equifax to {{Pay}} at {{Least}} \$650 {{Million}}
  in {{Largest}}-{{Ever Data Breach Settlement}}}.
\newblock \bibinfo{journal}{\emph{The New York Times}} (\bibinfo{year}{2019}).
\newblock


\bibitem[\protect\citeauthoryear{Dastin}{Dastin}{2018}]%
        {dastin2018AmazonScrapsSecret}
\bibfield{author}{\bibinfo{person}{Jeffrey Dastin}.}
  \bibinfo{year}{2018}\natexlab{}.
\newblock \showarticletitle{Amazon Scraps Secret {{AI}} Recruiting Tool That
  Showed Bias against Women}.
\newblock \bibinfo{journal}{\emph{Reuters}} (\bibinfo{year}{2018}).
\newblock


\bibitem[\protect\citeauthoryear{{David Anderson}, {Joy Bonaguro}, {Miriam
  McKinney}, and {Andrew Nicklin}}{{David Anderson} et~al\mbox{.}}{[n.d.]}]%
        {davidandersonEthicsAlgorithmsToolkit}
\bibfield{author}{\bibinfo{person}{{David Anderson}}, \bibinfo{person}{{Joy
  Bonaguro}}, \bibinfo{person}{{Miriam McKinney}}, {and}
  \bibinfo{person}{{Andrew Nicklin}}.} \bibinfo{year}{[n.d.]}\natexlab{}.
\newblock \bibinfo{title}{Ethics \& {{Algorithms Toolkit}} (Beta)}.
\newblock \bibinfo{howpublished}{https://ethicstoolkit.ai/}.
\newblock


\bibitem[\protect\citeauthoryear{Desmarais and Lowder}{Desmarais and
  Lowder}{2019}]%
        {desmarais2019PRETRIALRISKASSESSMENT}
\bibfield{author}{\bibinfo{person}{Sarah~L Desmarais} {and}
  \bibinfo{person}{Evan~M Lowder}.} \bibinfo{year}{2019}\natexlab{}.
\newblock \showarticletitle{{{PRETRIAL RISK ASSESSMENT TOOLS}}}.
\newblock  (\bibinfo{year}{2019}), \bibinfo{pages}{12}.
\newblock


\bibitem[\protect\citeauthoryear{DeVito, Birnholtz, Hancock, French, and
  Liu}{DeVito et~al\mbox{.}}{2018a}]%
        {devito2018people}
\bibfield{author}{\bibinfo{person}{Michael~A DeVito}, \bibinfo{person}{Jeremy
  Birnholtz}, \bibinfo{person}{Jeffery~T Hancock}, \bibinfo{person}{Megan
  French}, {and} \bibinfo{person}{Sunny Liu}.}
  \bibinfo{year}{2018}\natexlab{a}.
\newblock \showarticletitle{How people form folk theories of social media feeds
  and what it means for how we study self-presentation}. In
  \bibinfo{booktitle}{\emph{Proceedings of the 2018 CHI Conference on Human
  Factors in Computing Systems}}. ACM, \bibinfo{pages}{120}.
\newblock


\bibitem[\protect\citeauthoryear{DeVito, Hancock, French, Birnholtz, Antin,
  Karahalios, Tong, and Shklovski}{DeVito et~al\mbox{.}}{2018b}]%
        {devito2018algorithm}
\bibfield{author}{\bibinfo{person}{Michael~A DeVito},
  \bibinfo{person}{Jeffrey~T Hancock}, \bibinfo{person}{Megan French},
  \bibinfo{person}{Jeremy Birnholtz}, \bibinfo{person}{Judd Antin},
  \bibinfo{person}{Karrie Karahalios}, \bibinfo{person}{Stephanie Tong}, {and}
  \bibinfo{person}{Irina Shklovski}.} \bibinfo{year}{2018}\natexlab{b}.
\newblock \showarticletitle{The algorithm and the user: How can hci use lay
  understandings of algorithmic systems?}. In
  \bibinfo{booktitle}{\emph{Extended Abstracts of the 2018 CHI Conference on
  Human Factors in Computing Systems}}. ACM, \bibinfo{pages}{panel04}.
\newblock


\bibitem[\protect\citeauthoryear{Dillahunt, Erete, Galusca, Israni, Nacu, and
  Sengers}{Dillahunt et~al\mbox{.}}{2017}]%
        {dillahunt_reflections_2017}
\bibfield{author}{\bibinfo{person}{Tawanna~R. Dillahunt},
  \bibinfo{person}{Sheena Erete}, \bibinfo{person}{Roxana Galusca},
  \bibinfo{person}{Aarti Israni}, \bibinfo{person}{Denise Nacu}, {and}
  \bibinfo{person}{Phoebe Sengers}.} \bibinfo{year}{2017}\natexlab{}.
\newblock \showarticletitle{Reflections on {Design} {Methods} for {Underserved}
  {Communities}}. In \bibinfo{booktitle}{\emph{Companion of the 2017 {ACM}
  {Conference} on {Computer} {Supported} {Cooperative} {Work} and {Social}
  {Computing} - {CSCW} '17 {Companion}}}. \bibinfo{publisher}{ACM Press},
  \bibinfo{address}{Portland, Oregon, USA}, \bibinfo{pages}{409--413}.
\newblock
\showISBNx{978-1-4503-4688-7}
\urldef\tempurl%
\url{https://doi.org/10.1145/3022198.3022664}
\showDOI{\tempurl}


\bibitem[\protect\citeauthoryear{Donovan and Friedberg}{Donovan and
  Friedberg}{2019}]%
        {donovan2019source}
\bibfield{author}{\bibinfo{person}{Joan Donovan} {and} \bibinfo{person}{Brian
  Friedberg}.} \bibinfo{year}{2019}\natexlab{}.
\newblock \bibinfo{booktitle}{\emph{Source Hacking: Media Manipulation in
  Practice}}.
\newblock \bibinfo{type}{{T}echnical {R}eport}. \bibinfo{institution}{Data \&
  Society Research Institute}.
\newblock


\bibitem[\protect\citeauthoryear{Dressel and Farid}{Dressel and Farid}{2018}]%
        {dressel2018AccuracyFairnessLimits}
\bibfield{author}{\bibinfo{person}{Julia Dressel} {and} \bibinfo{person}{Hany
  Farid}.} \bibinfo{year}{2018}\natexlab{}.
\newblock \showarticletitle{The Accuracy, Fairness, and Limits of Predicting
  Recidivism}.
\newblock \bibinfo{journal}{\emph{Science Advances}} \bibinfo{volume}{4},
  \bibinfo{number}{1} (\bibinfo{year}{2018}).
\newblock


\bibitem[\protect\citeauthoryear{Edwards and Veale}{Edwards and Veale}{2017}]%
        {edwards2017SlaveAlgorithmWhy}
\bibfield{author}{\bibinfo{person}{Lilian Edwards} {and}
  \bibinfo{person}{Michael Veale}.} \bibinfo{year}{2017}\natexlab{}.
\newblock \showarticletitle{Slave to the {{Algorithm}}? {{Why}} a 'right to an
  Explanation' Is Probably Not the Remedy You Are Looking For}.
\newblock \bibinfo{journal}{\emph{Duke Law \& Technology Review}}
  \bibinfo{volume}{16} (\bibinfo{year}{2017}), \bibinfo{pages}{18--84}.
\newblock


\bibitem[\protect\citeauthoryear{Erete, Israni, and Dillahunt}{Erete
  et~al\mbox{.}}{2018}]%
        {erete_intersectional_2018}
\bibfield{author}{\bibinfo{person}{Sheena Erete}, \bibinfo{person}{Aarti
  Israni}, {and} \bibinfo{person}{Tawanna Dillahunt}.}
  \bibinfo{year}{2018}\natexlab{}.
\newblock \showarticletitle{An intersectional approach to designing in the
  margins}.
\newblock \bibinfo{journal}{\emph{Interactions}} \bibinfo{volume}{25},
  \bibinfo{number}{3} (\bibinfo{date}{April} \bibinfo{year}{2018}),
  \bibinfo{pages}{66--69}.
\newblock
\showISSN{10725520}
\urldef\tempurl%
\url{https://doi.org/10.1145/3194349}
\showDOI{\tempurl}


\bibitem[\protect\citeauthoryear{Eslami, Krishna~Kumaran, Sandvig, and
  Karahalios}{Eslami et~al\mbox{.}}{2018}]%
        {eslami_communicating_2018}
\bibfield{author}{\bibinfo{person}{Motahhare Eslami}, \bibinfo{person}{Sneha~R.
  Krishna~Kumaran}, \bibinfo{person}{Christian Sandvig}, {and}
  \bibinfo{person}{Karrie Karahalios}.} \bibinfo{year}{2018}\natexlab{}.
\newblock \showarticletitle{Communicating algorithmic process in online
  behavioral advertising}. In \bibinfo{booktitle}{\emph{Proceedings of the 2018
  {CHI} {Conference} on {Human} {Factors} in {Computing} {Systems}}}.
  \bibinfo{publisher}{ACM}, \bibinfo{pages}{432}.
\newblock


\bibitem[\protect\citeauthoryear{Eslami, Rickman, Vaccaro, Aleyasen, Vuong,
  Karahalios, Hamilton, and Sandvig}{Eslami et~al\mbox{.}}{2015}]%
        {eslami2015always}
\bibfield{author}{\bibinfo{person}{Motahhare Eslami}, \bibinfo{person}{Aimee
  Rickman}, \bibinfo{person}{Kristen Vaccaro}, \bibinfo{person}{Amirhossein
  Aleyasen}, \bibinfo{person}{Andy Vuong}, \bibinfo{person}{Karrie Karahalios},
  \bibinfo{person}{Kevin Hamilton}, {and} \bibinfo{person}{Christian Sandvig}.}
  \bibinfo{year}{2015}\natexlab{}.
\newblock \showarticletitle{I always assumed that I wasn't really that close to
  [her]: Reasoning about Invisible Algorithms in News Feeds}. In
  \bibinfo{booktitle}{\emph{Proceedings of the 33rd annual ACM conference on
  human factors in computing systems}}. ACM, \bibinfo{pages}{153--162}.
\newblock


\bibitem[\protect\citeauthoryear{Eslami, Vaccaro, Lee, Elazari Bar~On, Gilbert,
  and Karahalios}{Eslami et~al\mbox{.}}{2019}]%
        {eslami2019user}
\bibfield{author}{\bibinfo{person}{Motahhare Eslami}, \bibinfo{person}{Kristen
  Vaccaro}, \bibinfo{person}{Min~Kyung Lee}, \bibinfo{person}{Amit Elazari
  Bar~On}, \bibinfo{person}{Eric Gilbert}, {and} \bibinfo{person}{Karrie
  Karahalios}.} \bibinfo{year}{2019}\natexlab{}.
\newblock \showarticletitle{User Attitudes towards Algorithmic Opacity and
  Transparency in Online Reviewing Platforms}. In
  \bibinfo{booktitle}{\emph{Proceedings of the 2019 CHI Conference on Human
  Factors in Computing Systems}}. ACM, \bibinfo{pages}{494}.
\newblock


\bibitem[\protect\citeauthoryear{Gandal, Hamrick, Moore, and Oberman}{Gandal
  et~al\mbox{.}}{2018}]%
        {gandal2018price}
\bibfield{author}{\bibinfo{person}{Neil Gandal}, \bibinfo{person}{JT Hamrick},
  \bibinfo{person}{Tyler Moore}, {and} \bibinfo{person}{Tali Oberman}.}
  \bibinfo{year}{2018}\natexlab{}.
\newblock \showarticletitle{Price manipulation in the {Bitcoin} ecosystem}.
\newblock \bibinfo{journal}{\emph{Journal of Monetary Economics}}
  \bibinfo{volume}{95} (\bibinfo{year}{2018}), \bibinfo{pages}{86--96}.
\newblock


\bibitem[\protect\citeauthoryear{Gaver}{Gaver}{2012}]%
        {gaver2012WhatShouldWe}
\bibfield{author}{\bibinfo{person}{William Gaver}.}
  \bibinfo{year}{2012}\natexlab{}.
\newblock \showarticletitle{What Should We Expect from Research through
  Design?}. In \bibinfo{booktitle}{\emph{Proceedings of the 2012 {{ACM}} Annual
  Conference on {{Human Factors}} in {{Computing Systems}} - {{CHI}} '12}}.
  \bibinfo{publisher}{{ACM Press}}, \bibinfo{address}{{Austin, Texas, USA}},
  \bibinfo{pages}{937}.
\newblock


\bibitem[\protect\citeauthoryear{Gonz{\'a}lez}{Gonz{\'a}lez}{2017}]%
        {gonzalez2017SeattleSurveillanceOrdinance}
\bibfield{author}{\bibinfo{person}{M.~Lorena Gonz{\'a}lez}.}
  \bibinfo{year}{2017}\natexlab{}.
\newblock \bibinfo{title}{Seattle: {{Surveillance Ordinance}} ({{Seattle}})}.
\newblock
\newblock
\newblock
\shownote{Ordinance 125376.}


\bibitem[\protect\citeauthoryear{Green}{Green}{2018}]%
        {green_data_2018}
\bibfield{author}{\bibinfo{person}{Ben Green}.}
  \bibinfo{year}{2018}\natexlab{}.
\newblock \showarticletitle{Data {Science} as {Political} {Action}: {Grounding}
  {Data} {Science} in a {Politics} of {Justice}}.
\newblock  (\bibinfo{date}{Nov.} \bibinfo{year}{2018}).
\newblock
\urldef\tempurl%
\url{https://arxiv-org.offcampus.lib.washington.edu/abs/1811.03435v2}
\showURL{%
\tempurl}


\bibitem[\protect\citeauthoryear{Harrell}{Harrell}{2018}]%
        {harrell2018SeattleSurveillanceOrdinance}
\bibfield{author}{\bibinfo{person}{Bruce Harrell}.}
  \bibinfo{year}{2018}\natexlab{}.
\newblock \bibinfo{title}{Seattle: {{Surveillance Ordinance Amendment}}}.
\newblock
\newblock


\bibitem[\protect\citeauthoryear{Hoffmann}{Hoffmann}{2019}]%
        {hoffmann2019fairness}
\bibfield{author}{\bibinfo{person}{Anna~Lauren Hoffmann}.}
  \bibinfo{year}{2019}\natexlab{}.
\newblock \showarticletitle{Where fairness fails: data, algorithms, and the
  limits of antidiscrimination discourse}.
\newblock \bibinfo{journal}{\emph{Information, Communication \& Society}}
  \bibinfo{volume}{22}, \bibinfo{number}{7} (\bibinfo{year}{2019}),
  \bibinfo{pages}{900--915}.
\newblock


\bibitem[\protect\citeauthoryear{Inkpen, Chancellor, De~Choudhury, Veale, and
  Baumer}{Inkpen et~al\mbox{.}}{2019}]%
        {inkpen_where_2019}
\bibfield{author}{\bibinfo{person}{Kori Inkpen}, \bibinfo{person}{Stevie
  Chancellor}, \bibinfo{person}{Munmun De~Choudhury}, \bibinfo{person}{Michael
  Veale}, {and} \bibinfo{person}{Eric~PS Baumer}.}
  \bibinfo{year}{2019}\natexlab{}.
\newblock \showarticletitle{Where is the {Human}?: {Bridging} the {Gap}
  {Between} {AI} and {HCI}}. In \bibinfo{booktitle}{\emph{Extended {Abstracts}
  of the 2019 {CHI} {Conference} on {Human} {Factors} in {Computing}
  {Systems}}}. \bibinfo{publisher}{ACM}, \bibinfo{pages}{W09}.
\newblock


\bibitem[\protect\citeauthoryear{Koops, Newell, Timan, {\v S}korv{\'a}nek,
  Chokrevski, and Gali{\v c}}{Koops et~al\mbox{.}}{2016}]%
        {koops2016TypologyPrivacy}
\bibfield{author}{\bibinfo{person}{Bert-Jaap Koops},
  \bibinfo{person}{Bryce~Clayton Newell}, \bibinfo{person}{Tjerk Timan},
  \bibinfo{person}{Ivan {\v S}korv{\'a}nek}, \bibinfo{person}{Tom Chokrevski},
  {and} \bibinfo{person}{Ma{\v s}a Gali{\v c}}.}
  \bibinfo{year}{2016}\natexlab{}.
\newblock \showarticletitle{A Typology of Privacy}.
\newblock  (\bibinfo{year}{2016}).
\newblock
\newblock
\shownote{00012.}


\bibitem[\protect\citeauthoryear{Krafft, Della~Penna, and Pentland}{Krafft
  et~al\mbox{.}}{2018}]%
        {krafft2018experimental}
\bibfield{author}{\bibinfo{person}{P~M Krafft}, \bibinfo{person}{Nicol{\'a}s
  Della~Penna}, {and} \bibinfo{person}{Alex~Sandy Pentland}.}
  \bibinfo{year}{2018}\natexlab{}.
\newblock \showarticletitle{An experimental study of cryptocurrency market
  dynamics}. In \bibinfo{booktitle}{\emph{Proceedings of the 2018 CHI
  Conference on Human Factors in Computing Systems}}. ACM,
  \bibinfo{pages}{605}.
\newblock


\bibitem[\protect\citeauthoryear{Lecher}{Lecher}{2019}]%
        {lecher2019PrivacyAdvocateHeld}
\bibfield{author}{\bibinfo{person}{Colin Lecher}.}
  \bibinfo{year}{2019}\natexlab{}.
\newblock \showarticletitle{Privacy Advocate Held at Gunpoint after License
  Plate Reader Database Mistake, Lawsuit Alleges}.
\newblock \bibinfo{journal}{\emph{The Verge}} (\bibinfo{year}{2019}).
\newblock


\bibitem[\protect\citeauthoryear{Miller}{Miller}{2019}]%
        {miller2019explanation}
\bibfield{author}{\bibinfo{person}{Tim Miller}.}
  \bibinfo{year}{2019}\natexlab{}.
\newblock \showarticletitle{Explanation in artificial intelligence: Insights
  from the social sciences}.
\newblock \bibinfo{journal}{\emph{Artificial Intelligence}}
  \bibinfo{volume}{267} (\bibinfo{year}{2019}), \bibinfo{pages}{1--38}.
\newblock


\bibitem[\protect\citeauthoryear{Nagy and Neff}{Nagy and Neff}{2015}]%
        {nagy2015imagined}
\bibfield{author}{\bibinfo{person}{Peter Nagy} {and} \bibinfo{person}{Gina
  Neff}.} \bibinfo{year}{2015}\natexlab{}.
\newblock \showarticletitle{Imagined affordance: Reconstructing a keyword for
  communication theory}.
\newblock \bibinfo{journal}{\emph{Social Media+ Society}} \bibinfo{volume}{1},
  \bibinfo{number}{2} (\bibinfo{year}{2015}),
  \bibinfo{pages}{2056305115603385}.
\newblock


\bibitem[\protect\citeauthoryear{Rader, Cotter, and Cho}{Rader
  et~al\mbox{.}}{2018}]%
        {rader2018explanations}
\bibfield{author}{\bibinfo{person}{Emilee Rader}, \bibinfo{person}{Kelley
  Cotter}, {and} \bibinfo{person}{Janghee Cho}.}
  \bibinfo{year}{2018}\natexlab{}.
\newblock \showarticletitle{Explanations as mechanisms for supporting
  algorithmic transparency}. In \bibinfo{booktitle}{\emph{Proceedings of the
  2018 CHI Conference on Human Factors in Computing Systems}}. ACM,
  \bibinfo{pages}{103}.
\newblock


\bibitem[\protect\citeauthoryear{Rader and Gray}{Rader and Gray}{2015}]%
        {rader2015understanding}
\bibfield{author}{\bibinfo{person}{Emilee Rader} {and} \bibinfo{person}{Rebecca
  Gray}.} \bibinfo{year}{2015}\natexlab{}.
\newblock \showarticletitle{Understanding user beliefs about algorithmic
  curation in the Facebook news feed}. In \bibinfo{booktitle}{\emph{Proceedings
  of the 33rd annual ACM conference on human factors in computing systems}}.
  ACM, \bibinfo{pages}{173--182}.
\newblock


\bibitem[\protect\citeauthoryear{Rainie and Anderson}{Rainie and
  Anderson}{2017}]%
        {rainie2017theme}
\bibfield{author}{\bibinfo{person}{Lee Rainie} {and} \bibinfo{person}{Janna
  Anderson}.} \bibinfo{year}{2017}\natexlab{}.
\newblock \bibinfo{title}{The Need Grows for Algorithmic Literacy, Transparency
  and Oversight}.
\newblock
\newblock


\bibitem[\protect\citeauthoryear{Rosenblat and Stark}{Rosenblat and
  Stark}{2016}]%
        {rosenblat2016AlgorithmicLaborInformation}
\bibfield{author}{\bibinfo{person}{A. Rosenblat} {and} \bibinfo{person}{L.
  Stark}.} \bibinfo{year}{2016}\natexlab{}.
\newblock \showarticletitle{Algorithmic Labor and Information Asymmetries:
  {{A}} Case Study of {{Uber}}'s Drivers}.
\newblock \bibinfo{journal}{\emph{International Journal of Communication}}
  \bibinfo{volume}{10} (\bibinfo{year}{2016}), \bibinfo{pages}{3758--3784}.
\newblock


\bibitem[\protect\citeauthoryear{Selbst, Boyd, Friedler, Venkatasubramanian,
  and Vertesi}{Selbst et~al\mbox{.}}{2019}]%
        {selbst2019fairness}
\bibfield{author}{\bibinfo{person}{Andrew~D Selbst}, \bibinfo{person}{Danah
  Boyd}, \bibinfo{person}{Sorelle~A Friedler}, \bibinfo{person}{Suresh
  Venkatasubramanian}, {and} \bibinfo{person}{Janet Vertesi}.}
  \bibinfo{year}{2019}\natexlab{}.
\newblock \showarticletitle{Fairness and abstraction in sociotechnical
  systems}. In \bibinfo{booktitle}{\emph{Proceedings of the Conference on
  Fairness, Accountability, and Transparency}}. ACM, \bibinfo{pages}{59--68}.
\newblock


\bibitem[\protect\citeauthoryear{Selbst and Powles}{Selbst and Powles}{2017}]%
        {selbst2017meaningful}
\bibfield{author}{\bibinfo{person}{Andrew~D Selbst} {and}
  \bibinfo{person}{Julia Powles}.} \bibinfo{year}{2017}\natexlab{}.
\newblock \showarticletitle{Meaningful information and the right to
  explanation}.
\newblock \bibinfo{journal}{\emph{International Data Privacy Law}}
  \bibinfo{volume}{7}, \bibinfo{number}{4} (\bibinfo{year}{2017}),
  \bibinfo{pages}{233--242}.
\newblock


\bibitem[\protect\citeauthoryear{Star}{Star}{1999}]%
        {star1999ethnography}
\bibfield{author}{\bibinfo{person}{Susan~Leigh Star}.}
  \bibinfo{year}{1999}\natexlab{}.
\newblock \showarticletitle{The ethnography of infrastructure}.
\newblock \bibinfo{journal}{\emph{American behavioral scientist}}
  \bibinfo{volume}{43}, \bibinfo{number}{3} (\bibinfo{year}{1999}),
  \bibinfo{pages}{377--391}.
\newblock


\bibitem[\protect\citeauthoryear{Sturgis and Allum}{Sturgis and Allum}{2004}]%
        {sturgis2004science}
\bibfield{author}{\bibinfo{person}{Patrick Sturgis} {and} \bibinfo{person}{Nick
  Allum}.} \bibinfo{year}{2004}\natexlab{}.
\newblock \showarticletitle{Science in society: re-evaluating the deficit model
  of public attitudes}.
\newblock \bibinfo{journal}{\emph{Public understanding of science}}
  \bibinfo{volume}{13}, \bibinfo{number}{1} (\bibinfo{year}{2004}),
  \bibinfo{pages}{55--74}.
\newblock


\bibitem[\protect\citeauthoryear{Tucker, Guess, Barber{\'a}, Vaccari, Siegel,
  Sanovich, Stukal, and Nyhan}{Tucker et~al\mbox{.}}{2018}]%
        {tucker2018social}
\bibfield{author}{\bibinfo{person}{Joshua~A Tucker}, \bibinfo{person}{Andrew
  Guess}, \bibinfo{person}{Pablo Barber{\'a}}, \bibinfo{person}{Cristian
  Vaccari}, \bibinfo{person}{Alexandra Siegel}, \bibinfo{person}{Sergey
  Sanovich}, \bibinfo{person}{Denis Stukal}, {and} \bibinfo{person}{Brendan
  Nyhan}.} \bibinfo{year}{2018}\natexlab{}.
\newblock \bibinfo{title}{Social media, political polarization, and political
  disinformation: A review of the scientific literature}.
\newblock
\newblock


\bibitem[\protect\citeauthoryear{Wachter, Mittelstadt, and Russell}{Wachter
  et~al\mbox{.}}{2017}]%
        {wachter_counterfactual_2017}
\bibfield{author}{\bibinfo{person}{Sandra Wachter}, \bibinfo{person}{Brent
  Mittelstadt}, {and} \bibinfo{person}{Chris Russell}.}
  \bibinfo{year}{2017}\natexlab{}.
\newblock \showarticletitle{Counterfactual {Explanations} without {Opening} the
  {Black} {Box}: {Automated} {Decisions} and the {GPDR}}.
\newblock \bibinfo{journal}{\emph{Harv. JL \& Tech.}}  \bibinfo{volume}{31}
  (\bibinfo{year}{2017}), \bibinfo{pages}{841}.
\newblock


\bibitem[\protect\citeauthoryear{Woodruff, Fox, Rousso-Schindler, and
  Warshaw}{Woodruff et~al\mbox{.}}{2018}]%
        {woodruff2018qualitative}
\bibfield{author}{\bibinfo{person}{Allison Woodruff}, \bibinfo{person}{Sarah~E
  Fox}, \bibinfo{person}{Steven Rousso-Schindler}, {and}
  \bibinfo{person}{Jeffrey Warshaw}.} \bibinfo{year}{2018}\natexlab{}.
\newblock \showarticletitle{A qualitative exploration of perceptions of
  algorithmic fairness}. In \bibinfo{booktitle}{\emph{Proceedings of the 2018
  CHI Conference on Human Factors in Computing Systems}}. ACM,
  \bibinfo{pages}{656}.
\newblock


\bibitem[\protect\citeauthoryear{Woolley and Howard}{Woolley and
  Howard}{2018}]%
        {woolley2018computational}
\bibfield{author}{\bibinfo{person}{Samuel~C Woolley} {and}
  \bibinfo{person}{Philip~N Howard}.} \bibinfo{year}{2018}\natexlab{}.
\newblock \bibinfo{booktitle}{\emph{Computational Propaganda: Political
  Parties, Politicians, and Political Manipulation on Social Media}}.
\newblock \bibinfo{publisher}{Oxford University Press}.
\newblock


\bibitem[\protect\citeauthoryear{Young, Katell, and Krafft}{Young
  et~al\mbox{.}}{2019a}]%
        {young2019municipal}
\bibfield{author}{\bibinfo{person}{Meg Young}, \bibinfo{person}{Michael
  Katell}, {and} \bibinfo{person}{PM Krafft}.}
  \bibinfo{year}{2019}\natexlab{a}.
\newblock \showarticletitle{Municipal Surveillance Regulation and Algorithmic
  Accountability}.
\newblock \bibinfo{journal}{\emph{Big Data \& Society, Forthcoming}}
  (\bibinfo{year}{2019}).
\newblock


\bibitem[\protect\citeauthoryear{Young, Magassa, and Friedman}{Young
  et~al\mbox{.}}{2019b}]%
        {young2019InclusiveTechPolicy}
\bibfield{author}{\bibinfo{person}{Meg Young}, \bibinfo{person}{Lassana
  Magassa}, {and} \bibinfo{person}{Batya Friedman}.}
  \bibinfo{year}{2019}\natexlab{b}.
\newblock \showarticletitle{Toward Inclusive Tech Policy Design: A Method for
  Underrepresented Voices to Strengthen Tech Policy Documents}.
\newblock \bibinfo{journal}{\emph{Ethics and Information Technology}}
  \bibinfo{volume}{21}, \bibinfo{number}{2} (\bibinfo{year}{2019}),
  \bibinfo{pages}{89--103}.
\newblock


\bibitem[\protect\citeauthoryear{Zimmerman, Forlizzi, and Evenson}{Zimmerman
  et~al\mbox{.}}{2007}]%
        {zimmerman2007ResearchDesignMethoda}
\bibfield{author}{\bibinfo{person}{John Zimmerman}, \bibinfo{person}{Jodi
  Forlizzi}, {and} \bibinfo{person}{Shelley Evenson}.}
  \bibinfo{year}{2007}\natexlab{}.
\newblock \showarticletitle{Research through Design as a Method for Interaction
  Design Research in {{HCI}}}. In \bibinfo{booktitle}{\emph{Proceedings of the
  {{SIGCHI}} Conference on {{Human}} Factors in Computing Systems - {{CHI}}
  '07}}. \bibinfo{publisher}{{ACM Press}}, \bibinfo{address}{{San Jose,
  California, USA}}, \bibinfo{pages}{493}.
\newblock


\bibitem[\protect\citeauthoryear{Zimmerman, Stolterman, and Forlizzi}{Zimmerman
  et~al\mbox{.}}{[n.d.]}]%
        {zimmermanAnalysisCritiqueResearch}
\bibfield{author}{\bibinfo{person}{John Zimmerman}, \bibinfo{person}{Erik
  Stolterman}, {and} \bibinfo{person}{Jodi Forlizzi}.}
  \bibinfo{year}{[n.d.]}\natexlab{}.
\newblock \showarticletitle{An {{Analysis}} and {{Critique}} of {{Research}}
  through {{Design}}: Towards a Formalization of a Research Approach}.
\newblock  (\bibinfo{year}{[n.\,d.]}), \bibinfo{pages}{10}.
\newblock


\end{thebibliography}

\end{document}